\documentclass[12pt,epsf]{article}
\usepackage{graphicx}
\usepackage{amssymb,amsmath}

\begin{document}
%%%%%%%%%%%%%%%%%%%%%%%%%%%%%%%%%%%%%%%%%%%

\def\a{\alpha}
\def\b{\beta}
\def\c{\varepsilon}
\def\d{\delta}
\def\e{\epsilon}
\def\f{\phi}
\def\g{\gamma}
\def\h{\theta}
\def\k{\kappa}
\def\l{\lambda}
\def\m{\mu}
\def\n{\nu}
\def\p{\psi}
\def\q{\partial}
\def\r{\rho}
\def\s{\sigma}
\def\t{\tau}
\def\u{\upsilon}
\def\v{\varphi}
\def\w{\omega}
\def\x{\xi}
\def\y{\eta}
\def\z{\zeta}
\def\D{\Delta}
\def\G{\Gamma}
\def\H{\Theta}
\def\L{\Lambda}
\def\F{\Phi}
\def\P{\Psi}
\def\S{\Sigma}
\def\BR{{\rm Br}}
\def\o{\over}
\def\beq{\begin{eqnarray}}
\def\eeq{\end{eqnarray}}
\newcommand{\nn}{\nonumber \\}
\newcommand{\gsim}{ \mathop{}_{\textstyle \sim}^{\textstyle >} }
\newcommand{\lsim}{ \mathop{}_{\textstyle \sim}^{\textstyle <} }
\newcommand{\vev}[1]{ \left\langle {#1} \right\rangle }
\newcommand{\bra}[1]{ \langle {#1} | }
\newcommand{\ket}[1]{ | {#1} \rangle }
\newcommand{\EV}{ {\rm eV} }
\newcommand{\KEV}{ {\rm keV} }
\newcommand{\MEV}{ {\rm MeV} }
\newcommand{\GEV}{ {\rm GeV} }
\newcommand{\TEV}{ {\rm TeV} }
\def\diag{\mathop{\rm diag}\nolimits}
\def\Spin{\mathop{\rm Spin}}
\def\SO{\mathop{\rm SO}}
\def\O{\mathop{\rm O}}
\def\SU{\mathop{\rm SU}}
\def\U{\mathop{\rm U}}
\def\Sp{\mathop{\rm Sp}}
\def\SL{\mathop{\rm SL}}
\def\tr{\mathop{\rm tr}}

%added by FT
\newcommand{\bear}{\begin{array}}  
\newcommand {\eear}{\end{array}}
\newcommand{\la}{\left\langle}  
\newcommand{\ra}{\right\rangle}
\newcommand{\non}{\nonumber}  
\newcommand{\ds}{\displaystyle}
\newcommand{\red}{\textcolor{red}}
\def\ubl{U(1)$_{\rm B-L}$}
\def\REF#1{(\ref{#1})}
\def\lrf#1#2{ \left(\frac{#1}{#2}\right)}
\def\lrfp#1#2#3{ \left(\frac{#1}{#2} \right)^{#3}}
\def\OG#1{ {\cal O}(#1){\rm\,GeV}}

%%%%%%%%%%%%%%%%%%%%%%%%%%%%%%%
%%%    remove the following commands when finalizing
%%%%%%%%%%%%%%%%%%%%%%%%%%%%%%%
\def\TODO#1{ {\bf ($\clubsuit$ #1 $\clubsuit$)} }
%%%%%%%%%%%%%%%%%%%%%%%%%%%%%%%
%%%%%%%%%%%%%%%%%%%%%%%%%%%%%%%

%%%%%%%%%%%%%%%%%%%%%%%%%%%%%%%%%%%%%%%%%%%%%%%%%%%%%%%%%%%%%%%%%%%%

\baselineskip 0.7cm

\begin{titlepage}

\begin{flushright}
UT-12-27 \\
IPMU-12-0152
\end{flushright}

\vskip 1.35cm
\begin{center} 
{\large \bf 
Enhanced Diphoton Signal of the Higgs Boson\\
and the Muon $g-2$ in Gauge Mediation Models
}
\vskip 1.2cm

{Ryosuke Sato$^{a,b}$, Kohsaku Tobioka$^{a,b}$ and  Norimi Yokozaki$^{b}$}

\vskip 0.4cm

{\it
$^a$Department of Physics, University of Tokyo,\\
Tokyo 113-0033, Japan\\
$^b$Kavli Institute for the Physics and Mathematics of the Universe (WPI),\\
Todai Institutes for Advanced Study, University of Tokyo,\\
Kashiwa 277-8583, Japan\\
}

\vskip 1.5cm

\abstract{
We study the diphoton signal of the Higgs boson in gauge mediated supersymmetry breaking models, which can explain both the Higgs boson mass of around 125 GeV and the result of the muon $g-2$ experiment. We consider two possible extensions of gauge mediation models: inclusion of vector-like matters, and a mixing between a messenger and the up-type Higgs. 
The large left-right mixing of staus is induced in both scenarios, resulting in the enhanced diphoton signal. We include a constraint from a charge breaking minimum, which is severe for the large left-right mixing of staus. The branching fraction of $h \to \gamma \gamma$ can be about $40\%$ larger than that of the Standard Model Higgs boson, in the region of parameter space where the Higgs boson mass of around 125 GeV and the muon $g-2$ are explained.
 }
\end{center}
\end{titlepage}

\setcounter{page}{2}

%%%%%%%%%%%%%%%%%%%%%%%%%%%%%%%%%%%%%
\section{Introduction}
The Standard Model (SM) Higgs boson-like particle with a mass of around 125 GeV, has been discovered at both ATLAS~\cite{ATLAS_Higgs1} and CMS~\cite{CMS_Higgs1} experiments.
Although their observations are almost consistent with the prediction of the SM, there is an indication that the Higgs coupling to diphoton is somewhat enhanced~\cite{CMS_ATLAS_gamma}.

This may be explained if there is a new light charged particle which couples to the Higgs boson.
In the Minimal Supersymmetric (SUSY) Standard Model (MSSM), 
it was pointed out that
a light stau loop diagram can enhance a branching fraction of diphoton channel
in the case of a large left-right mixing of staus \cite{Carena:2011aa, Carena:2012gp}.
This possibility is interesting; the smuon is also light if soft masses are flavor-universal,
and the large left-right mixing of the stau is achieved by large $\tan\b$.
This is advantageous for the enhancement of 
the SUSY particle contribution to the muon anomalous magnetic moment (muon $g-2$) \cite{gminus2_SUSY}.
%On the other hand, it is known that 
%the SUSY particle contribution to the muon anomalous magnetic moment (muon $g-2$) is sizable in the case of light sleptons and large $\tan\b$ \cite{gminus2_SUSY}.
Actually, it was pointed out that
%assuming flavor-universal soft masses of sleptons,
there is  a possible correlation between the diphoton branching fraction of the Higgs boson and the SUSY contributions to muon $g-2$~\cite{Carena:2012gp,Giudice:2012pf}.
In fact, the experimental value of the muon $g-2$~\cite{Bennett:2006fi} is deviated from the prediction of the SM by $3.2\,\s$~\cite{Hagiwara:2011af} ($3.6\,\s$ \cite{dav}),
\begin{eqnarray}
a_\mu^{\rm exp}-a_\mu^{\rm SM}= (26.1 \pm 8.0) \times 10^{-10}. \label{eq:gm2}
\end{eqnarray}
Therefore, the light sleptons with large $\tan\b$ can explain the enhancement of diphoton signal and the deviation of the muon $g-2$ simultaneously.
However, in this case,
it is rather non-trivial whether a SUSY breaking scenario can also explain the Higgs boson mass of around 125 GeV.
For instance, it is difficult to explain the Higgs mass and the muon $g-2$ simultaneously in a mSUGRA scenario~\cite{msugra_dame}.

In this letter, we show that the diphoton signal of the Higgs boson can be enhanced in two gauge mediation models. These models can explain the Higgs boson mass of around 125 GeV and the muon $g-2$, simultaneously. The first model is a gauge mediation model with vector-like matters~\cite{gmsbvec1, gmsbvec2, Nakayama:2012zc,Martin:2012dg}, and the second one is a model which has a mixing between a messenger and the up-type Higgs, generating a large trilinear coupling of stops~\cite{EIY}. 
We consider the constraint from the meta-stability of the electroweak symmetry breaking minimum~\cite{Rattazzi:1996fb, Hisano:2010re} (see also \cite{vac_recent} for recent discussion), which is important when the left-right mixing of the stau is large. Including this constraint, the relative size of the branching fraction ${\rm Br}(h \to \gamma \gamma)/{\rm Br}(h \to \gamma \gamma)_{\rm SM}$ can be enhanced about $40\%$.

%%%%%%%%%%%%%%%%%%%%%%%%%%%%%%%%%%%%%
\section{Gauge mediation model with vector-like matters}
The gauge mediation model with vector-like matters at the TeV scale is studied in Refs.~\cite{gmsbvec1,gmsbvec2,Nakayama:2012zc,Martin:2012dg}.
We introduce complete multiplets of $SU(5)$ as ${\bf 10}=(Q', \bar{U}', \bar{E}')$ and ${\bf \overline{10}}=(\bar{Q}', {U}', {E}')$,
and the superpotential is given by
\begin{eqnarray}
W = W_{\rm MSSM} + y' Q' H_u \bar{U}' + y'' \bar{Q}' H_d U' + M_{Q} Q' \bar{Q}' + M_{U} U' \bar{U}' + M_E E'\bar{E}'.
\end{eqnarray}
As shown in Refs. \cite{Nakayama:2012zc}, the SUSY invariant mass $M_Q, M_U, M_E$ as well as the Higgsino mass parameter $\m$
can be related to Peccei-Quinn (PQ) symmetry breaking scale.
Then, we can consider vector-like matters have the mass of TeV scale naturally.
The new interaction, $y' Q' H_u \bar{U}'$ enhances the Higgs mass radiatively, in a similar manner of top-stop loops for $y' \simeq 1$ and $M_{Q}\simeq M_{U} \sim 1$ TeV~\cite{Moroi:1991mg, Babu, Martin:2009bg, Asano:2011zt}.
In fact, $y' \simeq 1$ is natural, since $y'$ has a quasi infrared fixed point behaviour at the electroweak scale as $y' \simeq 1$, while $y''$ does not~\cite{Martin:2009bg}. Since $y''$ decreases the Higgs boson mass, we assume that $y''$ is negligibly small in the following analysis.~\footnote{
The suppression of $y''$ can be explained by PQ symmetry~\cite{Nakayama:2012zc}.
} 

Because of $y' \simeq 1$, the soft mass of the up-type Higgs, $m_{H_u}$, gets non-negligible radiative corrections from extra squarks as 
\begin{eqnarray}
\Delta m_{H_u}^2 \sim -\frac{3 y'^2}{8\pi^2} (2  m_{\tilde{q}}^2 ) \log \frac{M_{\rm mess}}{m_{\rm SUSY}}, \label{eq:dmhu2}
\end{eqnarray}
where $M_{\rm mess}$ is the messenger scale, and $m_{\tilde{q}}$ is the squark mass. The stop mass scale is denoted as $m_{\rm SUSY}$.
The size of $\m$ parameter and the soft mass of the Higgs are related
by the following electroweak symmetry breaking condition:
\begin{eqnarray}
\frac{m_Z^2}{2} \simeq -\m^2 - \frac{\tan^2\b }{\tan^2\b-1 }m_{H_u}^2
+ \frac{1}{\tan^2 \b -1} m_{H_d}^2.
\end{eqnarray}
Because of Eq. (\ref{eq:dmhu2}), the size of $\mu^2$ is about twice as large as that of the gauge mediated SUSY breaking model without the extra-matters.
This induces a large left-right mixing of staus as
\begin{eqnarray}
\mathcal{L} \simeq \frac{m_{\tau}}{v} \mu \tan\beta \tilde{L}_3 {H_u^{0}}^* \tilde{\tau}_R^* + h.c.,
\end{eqnarray}
where $\tilde{L}_3$ and $\tilde{\tau}_R$ are the left-handed stau and the right-handed stau, respectively.
Due to the large mixing, the lightest stau becomes the next-to-lightest SUSY particle (NLSP) for moderately large $\tan\beta$, even when the number of the messengers (${\bf 5}$ and ${\bf \bar{5}}$ multiplets) is $N_{\rm mess}=1$; the stau mass can be lighter than the lightest neutralino mass due to the large left-right mixing.

As shown in Ref.~\cite{Carena:2011aa, Carena:2012gp}, the light stau with the large mixing is welcome to enhance the decay rate of the Higgs boson into $\gamma \gamma$. The stau contribution to $h$$\gamma$$\gamma$ coupling is approximately proportional to $(m_{\tau} |\mu| \tan\beta(m_{\tilde{\tau}_1}^{-2}-m_{\tilde{\tau}_2}^{-2}))$ for equal soft masses for $\tilde{L}_3$ and $\tilde{\tau}_R$, and it is constructive to the contribution from $W$ boson. Then, the relative size of the branching fractions, $r_{\gamma \gamma} \equiv {\rm Br}(h \to \gamma \gamma)/{\rm Br}(h \to \gamma \gamma)_{\rm SM}   \simeq \G(h\to\g\g)/\G(h\to\g\g)_{SM} $ is larger for the larger left-right mixing and the smaller stau mass.
Here, we consider the case that the mass of the CP-odd Higgs, $m_A$,
is large as $m_A \gg m_Z$. 

On the other hand, the large stau mixing induces deep charge breaking global minimum, and the electroweak symmetry breaking minimum becomes meta-stable~\cite{Rattazzi:1996fb, Hisano:2010re}.
Requiring that the life-time of the electroweak symmetry breaking local minimum is longer than the age of the universe, the upper-bound of $\mu \tan\beta$ is given by~\cite{Hisano:2010re}~\footnote{In Ref.~\cite{Hisano:2010re}, the vacuum transition rate is evaluated at the zero temperature.}
\begin{eqnarray}
\mu\tan\beta &<& 213.5 \sqrt{m_{\tilde{L}_3} m_{\tilde{\tau}_R}} -17.0 (m_{\tilde{L}_3} + m_{\tilde{\tau}_R}) \nonumber \\
&& +4.52\times10^{-2}\, {\rm GeV}^{-1} (m_{\tilde{L}_3} - m_{\tilde{\tau}_R})^2 -1.30 \times 10^4\ {\rm GeV}, \label{eq:fitting}
\end{eqnarray}
where $m_{\tilde{L}_3}$ and $m_{\tilde{\tau}_R}$ are soft masses of the left-handed and the right-handed staus, respectively. The fitting formula, Eq.(\ref{eq:fitting}) is not sensitive to $m_h$ and $\tan\beta$ for fixed $\mu\tan\beta$. 
%The dependence of the $\tan\beta$ is about $5 \%$.

\begin{figure}[t]
\begin{center}
\includegraphics[scale=1.05]{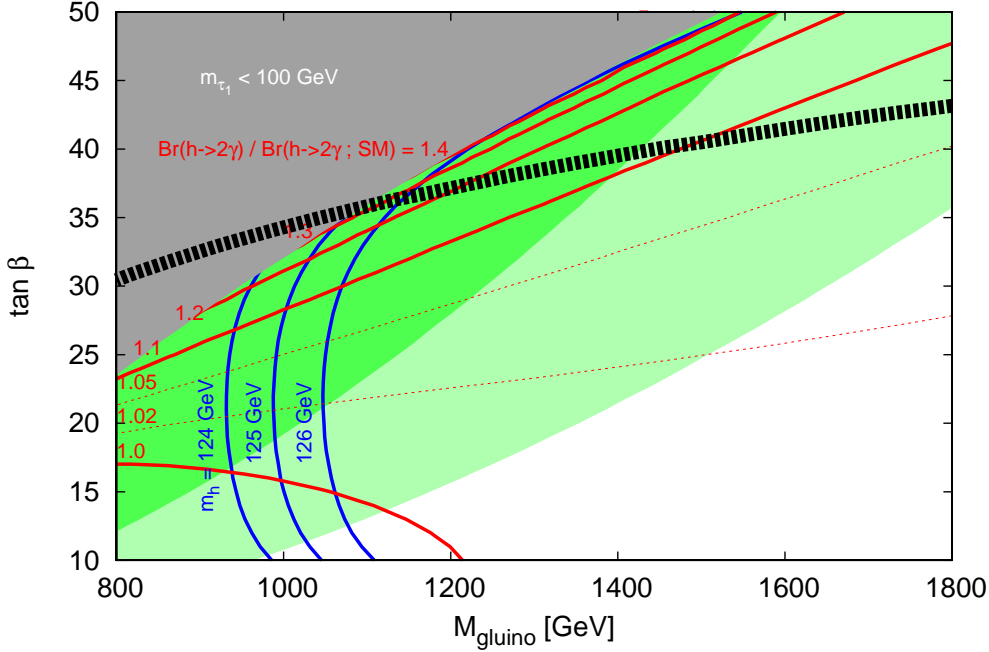}
\caption{Contours of $m_{h}$ and $r_{\gamma \gamma}$. The SUSY masses of the vector-like matters are set to $M_{Q}=M_{U}=700$ GeV. The messenger scale is set to $M_{\rm mess}=4 \times 10^5$ GeV. The other parameters are taken as $N_{\rm mess}=1$ and $y'(m_{\rm SUSY})=1.0$. Here, $m_t=173.2$ GeV and $\alpha_S(m_Z)=0.1184$. (The vacuum stability bound has changed, taking into account updated results in Ref.~\cite{Hisano:2010re})}
\label{fig:vec1}
\end{center}
\end{figure}

%%%%%%%%%%%%%%%%%% Explanation for figures
In Fig.~\ref{fig:vec1}, the contours of the Higgs mass and $r_{\gamma \gamma}$  are shown on $m_{\tilde{g}}$-$\tan\beta$ plane. We take the messenger number as $N_{\rm mess}=1$ and the messenger scale as $M_{\rm mess}= 4 \times 10^5$ GeV. The SUSY masses, $M_{Q}=M_{U}$, are taken as $700$ GeV.
The mass spectrum of the SUSY particles is calculated by using {\tt SuSpect 2.41}~\cite{suspect}, which is modified to include the renormalization group equations from the vector-like matters at the two-loop level. The Higgs boson mass, $r_{\gamma\gamma}$ and the SUSY contribution to the muon $g-2$ are calculated by {\tt FeynHiggs 2.9.2}~\cite{FeynHiggs}.
In the region within the dark (light) green band, the muon $g-2$ is explained at 1$\sigma$ (2$\sigma$) level (see Eq. (\ref{eq:gm2})). The black dashed line corresponds to the upper-bound of $\tan\beta$, which comes from meta-stability of the electroweak symmetry breaking vacuum. In the region around $m_{\tilde{g}}\simeq 1$ TeV, $r_{\gamma \gamma}$ can reach $1.3$. The Higgs boson mass, $m_{h} \simeq 125$ GeV,  and the muon $g-2$ (1$\sigma$) are also explained, simultaneously.

\begin{figure}[t]
\begin{center}
\includegraphics[scale=1.05]{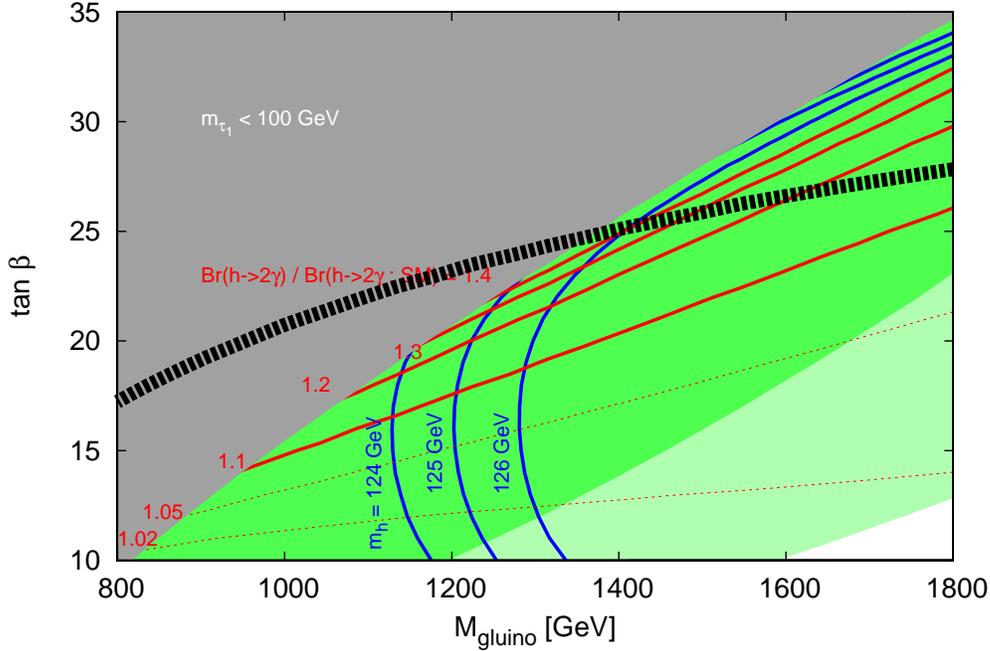}
\caption{Contours of $m_{h}$ and $r_{\gamma \gamma}$. The SUSY masses of the vector-like matters are set to $M_{Q}=M_{U}=900$ GeV. The SUSY breaking F-terms for $SU(2)_L$ doublet and $SU(3)_C$ triplet messengers are taken as $F_L/F_{\bar{D}}=0.4$. The messenger scale as well as the other parameters are same as in Fig.~\ref{fig:vec1}.}
\label{fig:vec2}
\end{center}
\end{figure}

As for the messenger fields, we have down-type messengers $\Phi_D,~\Phi_{\bar D}$
and lepton-type messengers $\Phi_{\bar L}, \Phi_L$, which are transformed as $(3,1)_{-1/3}, ({\bar 3},1)_{1/3}$,
and $(1,2)_{1/2}, (1,2)_{-1/2}$ under the SM gauge group, respectively.
The SUSY invariant mass terms and the SUSY breaking F-terms are given by
\begin{eqnarray}
W = (M_{\rm mess} + F_{\bar D} \theta^2 ) \Phi_D \Phi_{\bar D} + (M_{\rm mess} + F_L \theta^2 )\Phi_{\bar L} \Phi_L .
\end{eqnarray}
In Fig.~\ref{fig:vec1}, we assume that the SUSY breaking mass terms are universal, i.e., $F_{\bar D} = F_L$.~\footnote{To be more precise, we assume $F_L/M_{L}=F_{\bar{D}}/M_{\bar{D}}$.}
However, we may be able to split the SUSY breaking mass term.
In Fig.~\ref{fig:vec2}, we take $F_L  / F_{\bar D} = 0.4$, and $M_Q = M_U = 900~{\rm GeV}$.
It can be seen that, even if $m_{\tilde g} \simeq 1.4~{\rm TeV}$,
there is a parameter region which has the lightest Higgs mass around 125 GeV,
and can explain the muon $g-2$ at $1\s$ level. The diphoton ratio, $r_{\gamma \gamma}$, can reach about 1.4.

%%%%%%%%%%%%%%%%%%%%%%%%%%%%%%%%%%%%%%%%

\begin{figure}[t]
\begin{center}
\includegraphics[scale=1.05]{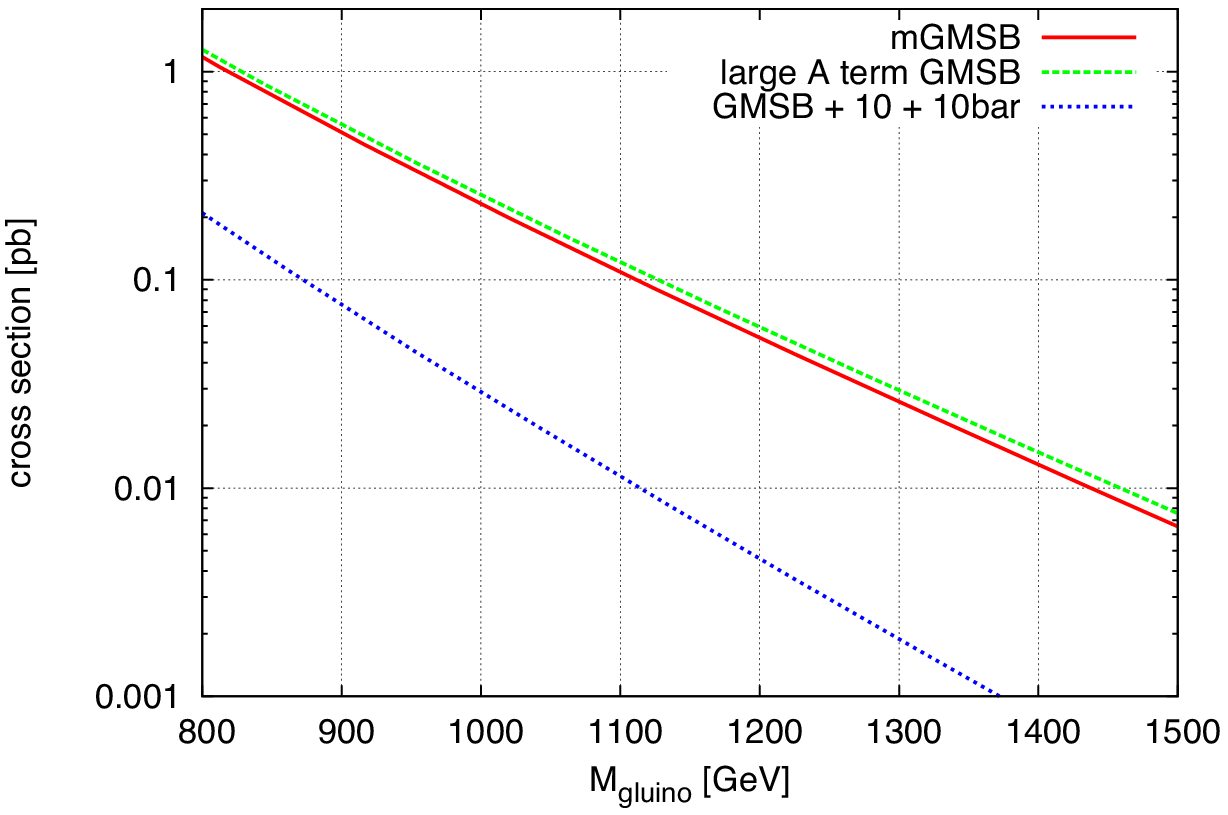}
\caption{The production cross sections of colored SUSY particles ($\tilde{g}\tilde{g} + \tilde{g}\tilde{q}+\tilde{q}\tilde{q} + \tilde{q}\tilde{q}^*$) are shown in different models: the minimal gauge mediation model with $N_{\rm mess} = 3$ and $M_{\rm mess}$ = $2.5 \times 10^5$ GeV (red solid line), the gauge mediation model generating a large $A_t$ with $N_{\rm mess} = 3$, $\tilde{y}_t = 1.4$ and $F_{\rm mess} / M_{\rm mess}^2 = 0.5$ (green dashed line), the gauge mediation model including the vector-like matters with $M_Q = M_U = 700$ GeV, $M_{\rm mess} = 4\times 10^5$ GeV and $y' = 1$ (blue dotted line). }
\label{fig:cs1}
\end{center}
\end{figure}

Let us comment on the constraint from the results of the SUSY search.
In the region with $r_{\rm \gamma \gamma} \gtrsim 1.1$, the stau is lighter than the neutralino, and thus the stau should decays into $\tau$ and gravitino (or other particles, e.g., axino) promptly, \footnote{
In the case of long-lived stau, its mass is severely constrained as $m_{\tilde{\tau}} \gtrsim 300$ GeV~\cite{ATLAS_2012_075}.
}
leaving tau(s), jets and missing transverse energy.
Such a channel is analyzed by both ATLAS \cite{ATLAS_tauplusjet_2012Aug} and CMS \cite{CMS_tauplusjet} experiments. It is shown that $m_{\tilde{q}} \simeq m_{\tilde{g}} \lesssim 1.2$ TeV, is excluded in the minimal gauge mediation model~\cite{ATLAS_tauplusjet_2012Aug}.
However, the production cross section of the colored SUSY particles are sensitive to their masses.  In the model with the vector-like matters, squark masses tend to be larger than the gluino mass. For example, $m_{\tilde{g}}\simeq 1$ TeV corresponds to $m_{\tilde{q}} \simeq 1.6$ TeV. Therefore the constraint on the gluino masses is expected to be less severe, due to the small production cross sections of colored SUSY particles. As a matter of fact, the production cross section evaluated by {\tt Prospino 2.1}~\cite{prospino} at the LHC 8 TeV is about 0.1 times smaller than the cross section in the reference point of Ref.~\cite{ATLAS_tauplusjet_2012Aug}, in the region of the relevant parameter space (see Fig.~{\ref{fig:cs1}}).
 
%The meta-stability bound is found in Ref.~\cite{Hisano:2010re}, which is evaluated at the zero temparature.~\footnote{The finite temperature effect may be important.}
%Note that the contributions to the gluon fusion process as well as $h \to \gamma \gamma$ from the vector-like matters are negligible for $M_{Q,U,E} \gtrsim ??$ GeV[].
%%%%%%%%%%%%%%%
%%%%%%%%%%%%%%%

%%%%%%%%%%%%%%%
%%%%%%%%%%%%%%%

%%%%%%%%%%%%%%%
%%%%%%%%%%%%%%%

%%%%%%%%%%%%%%%%%%%%%%%%%%%%%%%%%%%%%
\section{Gauge mediation model with large $A_t$}

The gauge mediation model with enhanced $A_t$ has been constructed in Refs.~\cite{EIY}.
A crucial point is the mixing between the up-type Higgs and the messenger. (A similar model was proposed in Refs.~\cite{Chacko:2001km} based on the framework of the extra dimension.) This mixing  induces a new interaction:
\begin{eqnarray}
W=y'_{U ij} \Phi_{\bar{L}} Q_i \bar{U}_j, \label{eq:EIY1}
\end{eqnarray}
where $\Phi_{\bar{L}}$ is the messenger, which has $(2,1/2)$ charge of $SU(2)_L \times U(1)_Y$.  The flavor dependent couplings, $y'_{U ij}$ are aligned to the up-quark Yukawa couplings of MSSM, $y_{u ij}$,  and therefore no dangerous flavor violating effects arise. After diagonalizing $y'_{U ij}$ and $y_{u ij}$, simultaneously,  Eq. (\ref{eq:EIY1}) is rewritten as 
\begin{eqnarray}
W=\tilde{y}_t \Phi_{\bar{L}} Q_3 \bar{U}_3, \label{eq:EIY2}
\end{eqnarray}
here we neglected the Yukawa couplings of the first and second generations. Due to the direct coupling between $\Phi_{\bar{L}}$ and top superfields, $A_t$ arises at one-loop level as
\begin{eqnarray}
A_t \simeq  -\frac{3 \tilde{y}_t^2}{16\pi^2}  \frac{F_{\rm mess}}{M_{\rm mess}}, \label{eq:dat}
\end{eqnarray}
where $F_{\rm mess}$ and $M_{\rm mess}$ are the SUSY breaking mass and the SUSY invariant mass of the messengers, respectively. As a matter of fact, the size of the generated $A_t$ is comparable with that of the stop masses generated by gauge interactions for $\tilde{y}_t \sim 1$. The additional contributions to soft masses of the stops and the up-type Higgs also arise as~\footnote{The soft masses of the sbottom and the down-type Higgs also get additional corrections, which are not so important here. (see Refs~\cite{EIY} for  complete formula.)}
\begin{eqnarray}
\delta_1 m_{Q_3}^2 &\simeq& \frac{\tilde{y}_t^2}{32\pi^2} \frac{F_{\rm mess}^2}{M_{\rm mess}^2} \left( 
\frac{(2+x)\log(1+x)+(2-x)\log(1-x)}{x^2}
\right),
\nonumber \\
\delta_1 m_{\bar{U}_3}^2 &\simeq& 2 \times \delta_1 m_{Q_3}^2, \label{eq:1loop}
\end{eqnarray}
at one loop level, and 
\begin{eqnarray}
\delta_2 m_{Q_3}^2 &\simeq& \frac{\tilde{y}_t^2}{128\pi^4} \frac{F_{\rm mess}^2}{M_{\rm mess}^2} \left( 3 \tilde{y}_t^2 + 3 y_t^2 -(8/3) g_3^2 -(3/2) g_2^2 -(13/30) g_1^2 \right), 
\nonumber \\
\delta_2 m_{\bar{U}_3}^2 &\simeq& \frac{\tilde{y}_t^2}{128\pi^4} \frac{F_{\rm mess}^2}{M_{\rm mess}^2} \left( 6 \tilde{y}_t^2 + 6 y_t^2 + y_b^2 -(16/3) g_3^2 -3 g_2^2 -(13/15) g_1^2 \right),\nonumber \\
\delta_2 m_{H_u}^2 &\simeq& -9\frac{\tilde{y}_t^2 y_t^2}{256\pi^4} \frac{F_{\rm mess}^2}{M_{\rm mess}^2}, \label{eq:2loop}
\end{eqnarray}
at two loop level. Here, we denote $y_t$ and $y_b$ as the top and bottom Yukawa coupling, respectively. The parameter $x$ in the one-loop contributions is defined as $x=F_{\rm mess}/M_{\rm mess}^2$. The two loop contribution can be calculated from wave-function renormalization~\cite{extracting_SUSY_breaking}, taking into account the kinetic mixing between $H_u$ and $\Phi_{\bar{L}}$ carefully (see Refs~\cite{EIY} for details). Note that  $\delta_1 m_{Q_3}^2$ and $\delta_1 m_{\bar{U}_3}^2$ vanish for $x \to 0$, while they are negative and comparable to $\delta_2 m_{Q_3}^2$ and $\delta_2 m_{\bar{U}_3}^2$ for $x \sim 1$. Since  the two-loop contributions are positive and large for $\tilde{y}_t \gtrsim 1$, the negative one-loop corrections are important to enlarge the ratio of $A_t/\sqrt{m_{Q_3} m_{\bar{U}_3}}$, which is crucial for the Higgs boson mass enhancement. As shown in Eq. (\ref{eq:2loop}), $m_{H_u}^2$ gets large negative corrections for $\tilde{y}_t \sim 1$, which is estimated as 
\begin{eqnarray}
\delta_2 m_{H_u}^2 \sim - (2\ {\rm  TeV})^2 \times \left[(100\ {\rm TeV})/(F_{\rm mess}/M_{\rm mess})\right]^2.
\end{eqnarray}
Therefore $\mu$ parameter is large in this model, resulting in the large left-right mixing of the staus.

The result of the numerical calculation is shown in Fig.~{\ref{fig:EIY1}}. In the calculation, the spectrum of the SUSY particles are calculated by using {\tt SuSpect 2.41}, which is modified to include the additional corrections, Eqs.\,({\ref{eq:dat}, {\ref{eq:1loop}, {\ref{eq:2loop}). As in the previous section, $m_h$, $r_{\gamma \gamma}$ as well as the SUSY contributions to the muon $g-2$ are calculated by using {\tt FeynHiggs 2.9.2}.
Here we assume that only a messenger mixes with $H_u$. We take the (total) number of messengers as $N_{\rm mess}=3$. 
We set $x=0.5$, and $\tilde{y}_t=1.4$. The enhanced rate, $r_{\gamma \gamma} \simeq 1.3$ is achieved with the Higgs mass of around $124$ GeV.~\footnote{There is an about $\pm 2$ GeV uncertainty in the calculation of the Higgs boson mass. In addition, $m_{h}$ can get about $+0.5$ GeV from three loop contributions for $m_{\tilde{t}} \simeq 1.2$ TeV~\cite{Martin:2007pg}.} The corresponding gluino mass is $m_{\tilde{g}} \simeq1.4$ TeV.
In the model with large $A_t$, the gluino mass is almost same as the squark masses, therefore the constraint in Ref.~\cite{ATLAS_tauplusjet_2012Aug} can apply to the gluino and the squark masses, that is, the region with $m_{\tilde{g}} \lesssim 1.2$ TeV is excluded.  
%%%%%%%%%%%%%%%
\begin{figure}[t]
\begin{center}
\includegraphics[scale=1.0]{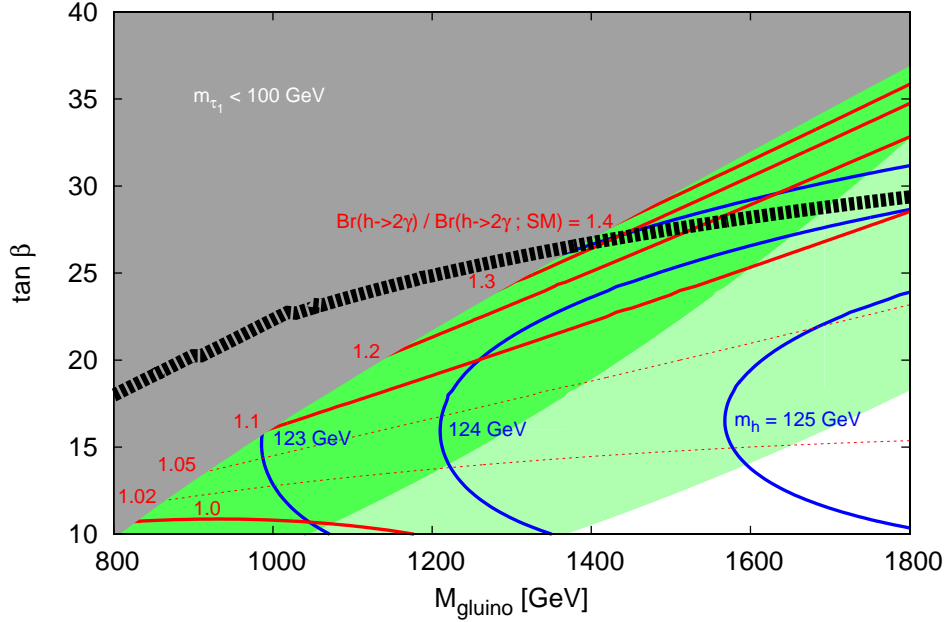}
\caption{Contours of $m_{h}$ and $r_{\gamma \gamma}$. The parameters are taken as $N_{\rm mess}=3$,  $x=0.5$, and $\tilde{y}_t=1.4$. Here, $m_t=173.2$ GeV and $\alpha_S(m_Z)=0.1184$.}
\label{fig:EIY1}
\end{center}
\end{figure}
%%%%%%%%%%%%%%%

%%%%%%%%%%%%%%%%%%%%%%%%%%%%%%%%%%%%%
\section{Conclusions and discussion}
We have evaluated the diphoton signal of the Higgs boson in two gauge mediated SUSY breaking models:  the model including vector-like matters, and the model having mixing between the up-type Higgs and the messenger.
 The large $\mu$-term, i.e., the large left-right mixing is induced automatically in both scenarios, and therefore the  enhanced diphoton signal of the Higgs boson is obtained. We have also considered the constraint from the charge breaking minimum of the staus, which gives the upper-bound to the size of the left-right mixing. The branching fraction of $h \to \gamma \gamma$ can be $1.4$ times larger than that of the SM Higgs boson. The Higgs boson mass of around 125 GeV and the muon $g-2$ are also explained in the both models.

The constraint from the charge breaking minimum is very important to scenarios which enhance the diphoton signal from stau loops, and thus it should be evaluated, including finite temperature effects.

\section*{Acknowledgements}
N.Y. thanks to Takeo Moroi, Kazunori Nakayama and Kyohei Mukaida for useful discussions.
The work of R.S, K.T. and N.Y. is supported in part by JSPS Research Fellowships for Young Scientists.
This work is also supported by the World Premier International Research Center Initiative (WPI Initiative), MEXT, Japan.

%\appendix

%\section{Vacuum stability in the large $B\mu$ case}
%The scalar potential for $H_u^0$, $H_d^0$, $\tilde{L}_3^-$ and $\tilde{\tau}_R^*$ is given by
%\begin{eqnarray}
%V &=& (m_{H_u}^2 + |\mu|^2) |H_u^0|^2 + (m_{H_d}^2 + |\mu|^2) |H_d^0|^2 - (B\mu H_u^0 H_d^0 + h.c.) \nonumber \\
%&+& m_{\tilde{L}_3}^2 |\tilde{L}_3^-|^2 + m_{\tilde{\tau}_R}^2 |\tilde{\tau}_R|^2-(y_{\tau} \mu \tilde{L}_3^- \tilde{\tau}_R^* {H_u^0}^* + h.c.)\nonumber \\
%&+& y_{\tau}^2 |\tilde{L}_3^- \tilde{\tau}_R^*|^2 + y_{\tau}^2 |H_d^0\tilde{L}_3^- |^2 + y_{\tau}^2 |H_d^0\tilde{\tau}_R^* |^2 \nonumber \\
%&+& \frac{g_2^2}{8}\left(|H_u^0|^2 + |\tilde{L}_3^-|^2 - |H_d^0|^2\right)^2 \nonumber \\
%&+& \frac{g_Y^2}{8}\left(|\tilde{L}_3^-|^2 - 2 |\tilde{\tau}_R|^2 -|H_u^0|^2+ |H_d^0|^2\right)^2,
%\end{eqnarray}
%where $g_2$ and $g_Y$ are gauge couplings of $SU(2)_L$ and $U(1)_Y$, respectively. Here, we neglect the A-term of the stau.

%\begin{itemize}
%\item D-flat direction
%\item EWSB at the origin
%\end{itemize}

%%%%%%%%%%%%%%%%%%%%%%%%%%%%%
%\bibliographystyle{JHEP}
%\bibliography{GMSB_PQ}

%%%%%%%%%%%%%%%%%%%%%%%%%%%%%%%%%%%%%

%%%%%%%%%%%%%%%%%%%%%%%%%%%%%%%%%%%%%
\end{document}